# Adsorption of CO on the Fe$_3$O$_4$(001) Surface

Jan Hulva[1], Zdeněk Jakub[1], Zbynek Novotny[1,2], Niclas Johansson[3], Jan Knudsen[3,4], Joachim Schnadt[3], Michael Schmid[1], Ulrike Diebold[1], Gareth S. Parkinson[1*]

[1] Institute of Applied Physics, TU Wien, Wiedner Hauptstraße 8-10, 1040, Vienna, Austria

[2] Present address: Department of Physics, University of Zürich, Winterthurerstrasse 190, 8057 Zürich, Switzerland

[3] Division of Synchrotron Radiation Research, Lund University, Box 118, SE-221 00 Lund, Sweden

[4] MAX IV Laboratory, Lund University, Box 118, SE-221 00 Lund, Sweden

[*] to whom correspondence should be addressed: parkinson@iap.tuwien.ac.at

**Abstract**

The interaction of CO with the Fe$_3$O$_4$(001)-($\sqrt{2}\times\sqrt{2}$)R45° surface was studied using temperature programmed desorption (TPD), scanning tunneling microscopy (STM) and x-ray photoelectron spectroscopy (XPS), the latter both under ultrahigh vacuum (UHV) conditions and in CO pressures up to 1 mbar. In general, the CO-Fe$_3$O$_4$ interaction is found to be weak. The strongest adsorption occurs at surface defects, leading to small TPD peaks at 115 K, 130 K and 190 K. Desorption from the regular surface occurs in two distinct regimes. For coverages up to 2 CO molecules per ($\sqrt{2}\times\sqrt{2}$)R45° unit cell, the desorption maximum shows a large shift with increasing coverage, from initially 105 K to 70 K. For coverages between 2 and 4 molecules per ($\sqrt{2}\times\sqrt{2}$)R45° unit cell, a much sharper desorption feature emerges at ≈50 K. Thermodynamic analysis of the TPD data suggests a phase transition from a dilute 2D gas into an ordered overlayer with CO molecules bound to surface Fe$^{3+}$ sites. XPS data acquired at 45 K in UHV are consistent with physisorption. Some carbon-containing species are observed in the near-ambient-pressure XPS experiments at room temperature, but are attributed to contamination and/or reaction with CO with water from the residual gas. No evidence was found for surface reduction or carburization by CO molecules.

## 1. Introduction

Iron oxides are omnipresent in the natural environment, and play a role in many industrial applications.[1] By far the greatest single use of iron oxides is as a source of iron ore for steel production, a major source of CO$_2$ emissions. In the first step, the mined minerals are heated in a CO-rich environment in a process called carbo-thermal reduction. A second important application is in catalysis; $Fe_3O_4$ is the active phase of the industrial high-temperature water-gas shift catalyst,[2-3] and is thought to play a role in Fe-based Fischer-Tropsch synthesis.[4-5] In both reactions, the adsorption of CO is an important precursor to any chemistry. It has been suggested that CO modifies the surface during the reactions, via reduction in the case of water-gas shift chemistry[2-3], and through carburization in Fischer-Tropsch synthesis[5]. In this paper, we study the interaction of CO with $Fe_3O_4$(001), with the additional motivation that CO adsorption is an often-used probe of the acid/base character of metal oxide surfaces. Strongly acidic sites result in chemisorption and/or the formation of carbonyls (for example on $Co_3O_4$(111)[6] and (001)[7], $Sr_3Ru_2O_7$(001)[8]), whereas surfaces with weakly acidic sites typically physisorb CO (e.g. $TiO_2$ rutile (110)[9-10] or anatase (101)[11]).

The existing literature regarding the interaction of CO with $Fe_3O_4$ is rather limited. Lemire et al.[12] used a combination of TPD and Infrared Reflection Absorption Spectroscopy (IRAS) to investigate the CO adsorption of $Fe_3O_4$(111) thin films grown on Pt (111). Three TPD peaks were observed at 230 K, 180 K and 110 K, and assigned to CO adsorbed at step edge $Fe^{3+}$ sites, $Fe^{2+}$ terrace sites, and highly-mobile physisorbed CO, respectively. The authors interpreted their data as evidence of an $Fe_{oct2}$ surface termination. Huang et al.[13] studied the same system using density functional theory (DFT), and predicted adsorption energies of 1.94 eV and 0.80 eV for the $Fe_{oct2}$ and $Fe_{tet2}$ terminations in the low coverage limit, respectively. These energies are surprisingly high, and likely do not correspond to the TPD peaks observed by Lemire et al.[12]. Nevertheless, the adsorption energies were predicted to decrease with coverage due to repulsive interaction between the CO molecules.

The only prior publication regarding CO adsorption on $Fe_3O_4$(001) is a DFT+U study by Xue et al.[14]. These authors predicted a much lower adsorption energy of 0.24 eV for an isolated CO molecule bound to a fivefold-coordinated surface $Fe^{3+}$ cation. Moreover, they propose that the adsorption energy of CO molecules should increase with coverage due to formation of C-C bonds, and that a direct oxidation of the adsorbed CO molecule to $CO_2$ is favorable. Unfortunately, the activation energies for these processes were not calculated. It is important to mention that the calculations were performed using a stoichiometric model for the (001) surface based on a truncation at the $Fe_{oct}$-O (B-layer) termination[14]. Recently, our group determined that the $(\sqrt{2}\times\sqrt{2})R45°$ reconstruction is based on an ordered array of cation vacancies and interstitials in the subsurface layers[15]. Hereafter, we refer to it as the subsurface cation vacancy (SCV) model. However, since the top surface layer does not differ much between the models, we do not expect a major difference in the adsorption energy for an isolated CO molecule.

In this paper we report an experimental study of the interaction of CO with the $Fe_3O_4(001)$-($\sqrt{2}\times\sqrt{2}$)R45° surface. Under ultra-high vacuum (UHV) conditions, CO binds weakly to the terrace sites, desorbing between 50 K and 105 K in TPD. The maximum coverage in the first monolayer is 4 molecules per ($\sqrt{2}\times\sqrt{2}$)R45° unit cell, which corresponds to the number of undercoordinated surface $Fe^{3+}$ cations in the surface layer. Repulsive interactions between adsorbed CO molecules lead to a strongly coverage-dependent shape of the TPD spectra, and a phase transition occurs close to 2 molecules per unit cell. CO molecules bind to surface defects more strongly than to the regular lattice sites, which is also observed by scanning tunneling microscopy (STM). Near ambient pressure X-ray Photoelectron Spectroscopy (NAP-XPS) up to 1 mbar reveal no significant difference to XPS taken under UHV conditions, and there is no evidence for surface reduction or carburization at room temperature. Exposure to the near ambient pressure environment does lead to carbonaceous deposits at the surface, but these most likely originate from contamination and/or reactions between CO and water in the residual gas.

## 2    Experimental Details

The UHV-XPS and TPD experiments were performed using a newly-constructed UHV setup optimized to study the surface chemistry of metal oxide single crystal samples[16]. The vacuum chamber achieves a base pressure of $8\times10^{-11}$ mbar, and is equipped with a range of facilities for surface spectroscopies and preparation. Pertinent here are a monochromatic Al Kα X-ray source, a SPECS Phoibos 150 hemispherical analyzer, a low-energy electron diffraction (LEED) optics, and a home-built calibrated molecular beam source. Full details of the setup can be found in Ref. [16]. A natural $Fe_3O_4(001)$ single crystal (Surface Preparation Laboratory) was mounted on a Ta sample holder using several Ta clips, and a thin gold foil was placed between the back of the sample and the sample plate to ensure good thermal contact. The sample plate was heated by direct current, and cooled by a liquid-He flow cryostat, and the temperature was measured by a K-type thermocouple spot-welded to its side.  The $Fe_3O_4(001)$ surface was prepared in-situ by cycles of $Ne^+$ sputtering followed by UHV annealing at 920 K. In the last cycle before the measurements the sample was annealed in an $O_2$ pressure of $1 \times 10^{-6}$ mbar resulting in a sharp ($\sqrt{2}\times\sqrt{2}$)R45° LEED pattern (not shown).

High-purity carbon monoxide (=99.997%) was dosed directly onto the sample surface using the effusive molecular beam source[17] at normal incidence. The beam is formed by expanding 0.53 mbar of CO gas (as measured by a capacitance gauge) through two differentially pumped stages, resulting in a well-defined beam spot with a diameter of 3.35±0.17 mm at the sample. The beam has a top-hat intensity profile and, based on the source pressure and the beam geometry, the flux is known to be $7.6 \pm 0.4 \times 10^{12}$ CO molecules/cm$^2$s.

The TPD experiments were performed with a heating rate of 1 K/s (unless otherwise stated) using a Hiden HAL 3F PIC mass spectrometer directly facing the sample with the mass spec axes aligned with the sample normal. Near the beginning of the temperature ramp the rate is nonlinear, which could be problematic as CO already begins to desorb from the surface at around 50 K for high coverages. Correction of the desorption rate for this nonlinearity had negligible influence on the parameters obtained from the TPD analysis. The time for the sample to cool down from 450 K to the dosing temperature of 45 K was ≈ 15 minutes.

Monochromatic XPS spectra (Al Kα) were measured at 45 K at grazing emission (80° from the surface normal). The energy scale was calibrated to the Fermi edge of the metal sample. Comparing the first and the last scan of each measurement shows a minor decrease (15 - 20 %) in the intensity of the CO-related peaks over the course of the experiment (data acquisition time ~ 30 min). This could be caused by the displacement of weakly adsorbed CO by more strongly bound molecules from the residual gas ($CO_2$, $H_2O$), or by a slow thermal or x-ray induced desorption. No further changes were observed in the spectra.

The STM experiments were performed at ≈ 80 K in a separate UHV system with a base pressure $5 \times 10^{-12}$ mbar using an Omicron LT-STM in constant-current mode with electrochemically etched W tips. In this case, a synthetic magnetite single crystal was prepared by 1 keV Ar$^+$ sputtering followed by heating to 920 K. Again, every other annealing cycle was performed in a background pressure of $1 \times 10^{-6}$ mbar $O_2$.

Near ambient pressure XPS measurements were conducted at the MAX-lab synchrotron in Lund, Sweden, using beamline I511 on the MAX II ring[18]. With this instrument, both UHV-based ($10^{-10}$ < p > $10^{-5}$ mbar) and NAP (near ambient pressure, $10^{-4}$ mbar < p < 10 mbar) measurements can be performed using the same SPECS PHOIBOS 150 NAP analyzer. Photon energies of 850 eV, 650 eV, and 390 eV were used to acquire the Fe *2p*, O *1s* and C *1s* regions, respectively, with a pass energy of 50 eV. The Fermi level was calibrated using the Fermi edge measured on the Mo sample plate.

# 3    Results

## 3.1    Sticking coefficient measurements

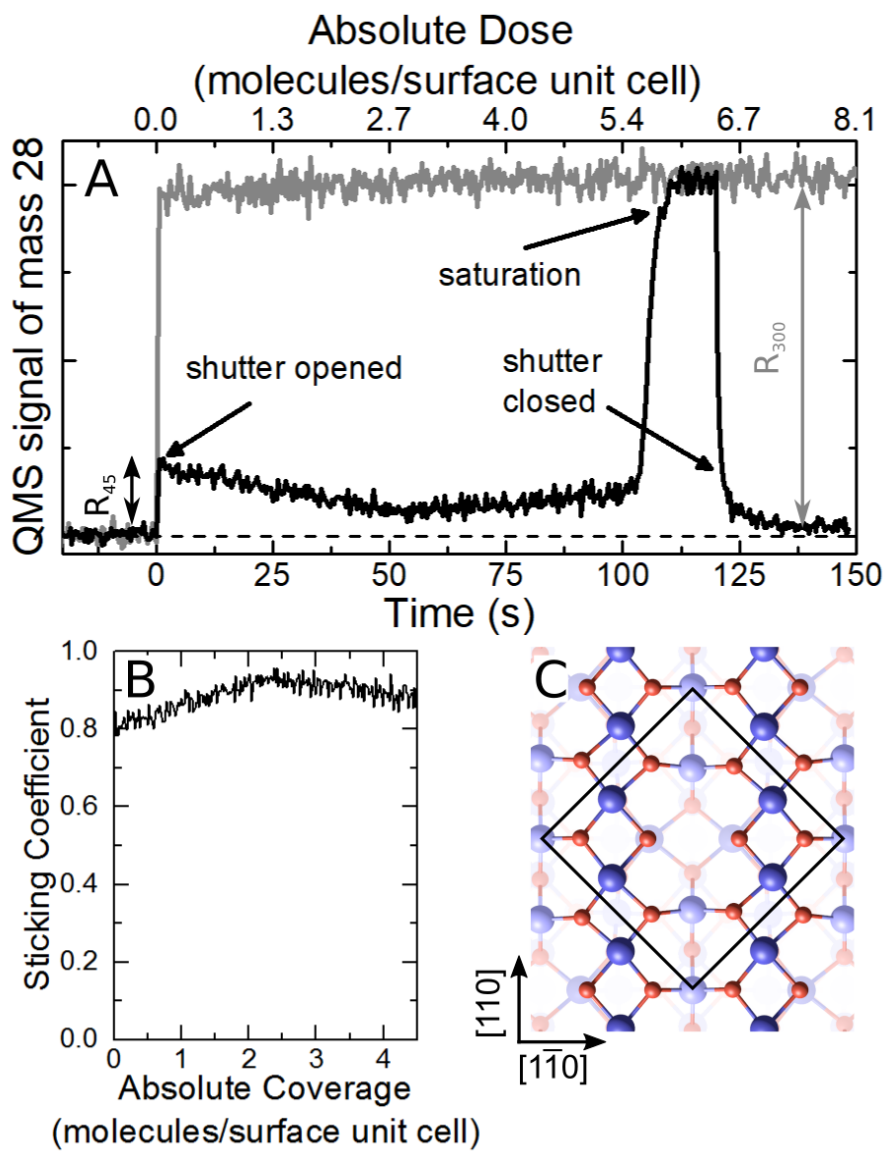

**Figure 1:** **(A)** Mass 28 signal (CO partial pressure) recorded during exposing a CO molecular beam on a $Fe_3O_4$(001) sample kept at 300 K (grey) and 45 K (black). After opening the beam shutter at 45 K, most of the CO molecules remain adsorbed on the cold sample until saturation occurs around a dose of 5.3 CO molecules/surface unit cell. **(B)** Sticking coefficient vs. coverage for CO dosed at 45 K. The reference for zero sticking was taken from the CO dosing at 300 K. **(C)** The SCV model for the $Fe_3O_4$ (001)-($\sqrt{2}\times\sqrt{2}$)R45° surface[15]. The ($\sqrt{2}\times\sqrt{2}$)R45° unit cell is indicated by the black square, and dark blue, light blue, and small red balls represent surface fivefold coordinated $Fe_{oct}$, subsurface fourfold coordinated $Fe_{tet}$ and oxygen atoms, respectively.

To determine the absolute coverage of CO molecules on the $Fe_3O_4$(001) surface requires both the beam flux and the sticking coefficient as a function of coverage at the dose temperature. Sticking coefficient measurements were performed using the well-known method of King and Wells[19], with the proviso that our experimental setup differs from the traditional experiment because the mass spectrometer is positioned at 45° from the surface normal (i.e., in a line-of-sight geometry). In the standard configuration, the sample surface is not line of sight with the mass spec, and one measures the spatially equilibrated increase of CO partial pressure. In our setup, we could measure a higher signal due to molecules that directly scatter from the sample surface into the mass spectrometer. This should not affect the absolute values of the sticking coefficient, since the zero-sticking reference was acquired in the same geometry. A variation in the angular distribution of the scattered molecules with coverage could influence our measurement, but as we demonstrate below, we do not think this is the case.

Figure 1A shows the CO mass spec intensity as a function of time and the corresponding CO dose, with the sample maintained at 300 K (grey curve) and 45 K (black curve). At 300 K, the CO signal reaches a maximum immediately after the opening the beam shutter and remains constant until the shutter is closed. This behavior is consistent with no CO adsorbing on the surface at 300 K. When CO is dosed at 45 K, the CO signal is significantly lower because much of it sticks on the sample. The signal decreases linearly with time until around 60 s (3.1 CO molecules per surface unit cell dosed at the sample). It then remains constant until a sharp step at 103 s (5.3 molecules per surface unit cell). The signal reaches the same intensity as observed at 300 K after 110 s (5.6 molecules per surface unit cell), which suggests no more CO can adsorb on the surface. CO multilayers are known to desorb around 30 K and thus cannot be condensed in this experiment[20]. After closing the beam shutter CO signal does not fall immediately to the background level because molecules desorb from the weakest bound states already at 45 K.

In Figure 1B we plot the sticking coefficient as a function of CO coverage. First, the sticking coefficient as a function of dosing time is calculated as

$$S(t) = \frac{R_{300}(t) - R_{45}(t)}{R_{300}(t)},$$

where $R_{45}$ is the (background subtracted) CO signal at 45 K, and $R_{300}$ is the CO signal at 300 K.

Next, the absolute coverage is calculated as

$$\theta_{abs} = I_{beam} t \int_0^t S(t') dt',$$

where $I_{beam}$ is the beam intensity ($7.6 \times 10^{12}$ CO molecules/cm²s)[16] and $t$ the dosing time.

Initially, the sticking coefficient is high ≈ 0.8, and increases with coverage up to ≈ 2 CO molecules per surface unit cell, where it reaches a maximum value of 0.93. The high sticking coefficient indicates a low barrier (if any) to adsorption[9], while the linear increase with coverage is attributed to the more efficient momentum transfer between incoming and adsorbed CO molecules compared to the bare $Fe_3O_4(001)$ substrate. Above 2 molecules per (√2×√2)R45° unit cell, the sticking coefficient remains constant until the saturation of the sample by CO at 45 K by ≈ 4.7 molecules per (√2×√2)R45° unit cell. High values of the sticking coefficient up to the saturation of the surface point to the existence of the mobile precursor state in the second layer.

## 3.2 Temperature-Programmed Desorption

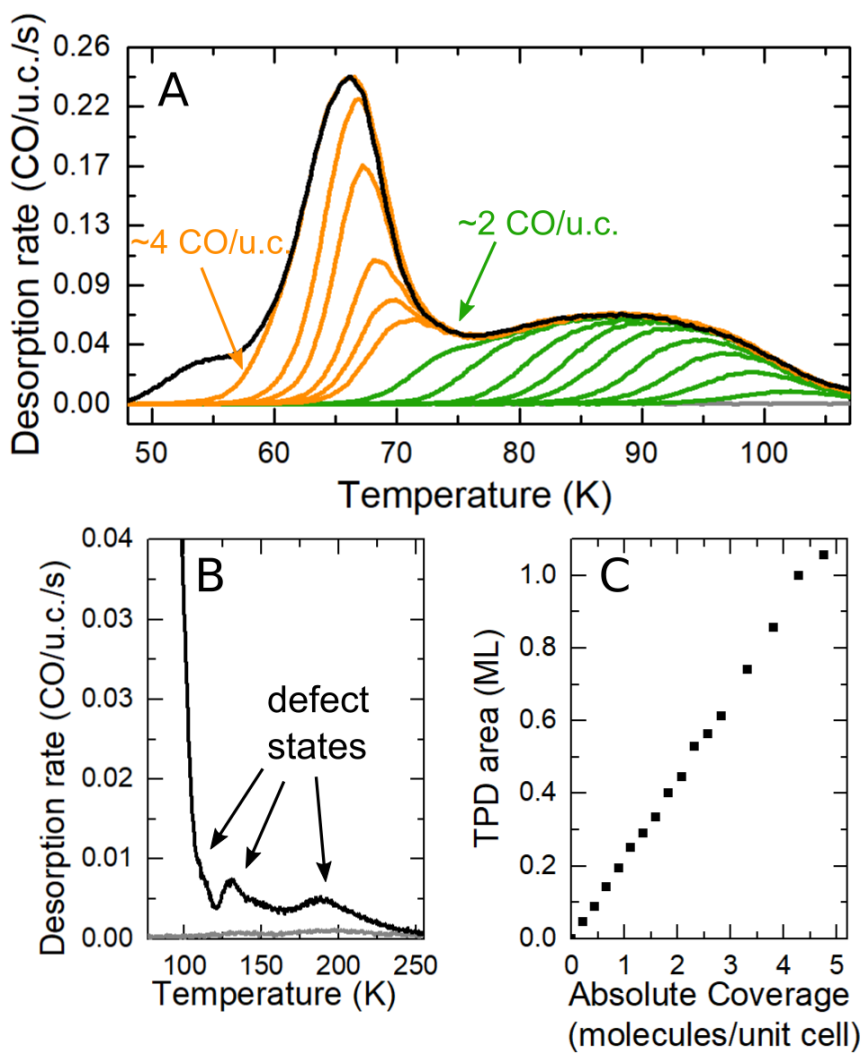

**Figure 2: (A)** TPD spectra for various initial coverages of CO (0 (grey), 0.07, 0.12, 0.17, 0.22, 0.28, 0.32, 0.37, 0.43, 0.48 (green), 0.56, 0.59, 0.64, 0.76, 0.87, 1.00 (orange), 1.04 (black) ML) with a heating rate of 1 K/s. 1 Monolayer (ML) is defined as the area of the saturated peak (last orange curve). **(B)** Detail of the higher temperature range showing low-intensity desorption states related to surface defects. The data is for an initial coverage of 1 ML. Grey curve shows background signal acquired without any CO dosing. **(C)** Plot of the coverage obtained by integrating the TPD curves versus the absolute coverage, determined from the calibrated beam dose and taking into account the measured the sticking coefficient. As expected, the relationship is linear.

In Figure 2A, we show TPD spectra acquired for a range of initial CO coverages with a heating rate of 1 K/s. The TPD spectra are normalized by the desorption curve corresponding to the dose of 1.8 L, which is taken as 1 monolayer (1 ML). At the lowest coverage measured (0.07 ML), CO desorbs in four peaks at 102 K (shoulder), 115 K, 130 K and 190 K (see also Fig. 2B). Given the low intensity of the three higher-temperature peaks, we attribute them to molecules adsorbed at surface defects. As we show below using STM (Section 3.3) these include antiphase domain boundaries (APDBs) in the ($\sqrt{2}\times\sqrt{2}$)R45° reconstruction[21], surface hydroxyls[22], Fe adatoms[23], incorporated Fe defects[23], and step edges. The total amount of CO desorbing from the surface defects is ≈5% of the monolayer coverage. We performed a series of experiments to investigate the origin of the surface defects (see figure S1 in supporting information). The broad feature between 130 K and 160 K seems to be correlated with the concentration of step-edges, whereas the sharp feature at 130 K is most likely related to $Fe^{2+}$ sites arising from surface hydroxylation and/or subsurface Fe defects. For higher coverages only the peak at ≈100 K increases in intensity. With increasing coverage the maximum shifts rapidly to lower temperatures (green curves in Fig. 2 A). Saturation of this broad feature is followed by the appearance of a much sharper peak at 65 K (orange curves in Fig. 2A). Defining the saturation of this sharp peak as 1 monolayer (ML) coverage, we find that the lower-coverage peak saturates at is ≈0.48 ML (green curves in Figure 2 A). Increasing the initial coverage above 1 ML results in the black curve (1.04 ML), which also includes a small shoulder at 55 K. This is the maximum amount of CO we can adsorb at the sample at 45 K. This feature is related to desorption from a compressed monolayer of CO molecules that typically forms before the multilayer peak in physisorbed systems [24] (see Figure S3 in supporting information).

The lower-coverage peak (0.48 ML, green traces in Figure 2 A) saturates at 2.1 ± 0.2 CO molecules per ($\sqrt{2}\times\sqrt{2}$)R45° unit cell, while the higher-coverage curve (defined as 1 ML, orange traces in Figure 2 A) saturates at a coverage of 4.3±0.4 CO molecules per ($\sqrt{2}\times\sqrt{2}$)R45° unit cell. This is straightforward to understand because there are 4 $Fe^{3+}$ cation sites available per ($\sqrt{2}\times\sqrt{2}$)R45° unit cell (Figure 1 C, dark blue atoms). This observation is in line with our previous study of $CO_2$, where saturation also occurred at 4 molecules per unit cell. Interestingly, this suggests that both peaks within the first monolayer (green and orange curves in Figure 2 A) originate from the same adsorption site.

Earlier, we mentioned the possibility that the line-of-sight geometry of our King and Wells experiment could present problems if the angular dependence of scattered CO molecules changed with coverage. To check this, we plot in Figure 2C the coverage determined by integration of the normalized TPD curves versus the absolute coverage determined by the sticking coefficient corrections of the beam intensity. The linear scaling shows that the CO coverage determination is consistent between adsorption/desorption methods across the range of coverages considered. (A minor deviation from the linear behavior at the highest coverage occurs because some CO molecules desorb between the dosing and the TPD measurement for saturation coverages at 45 K).

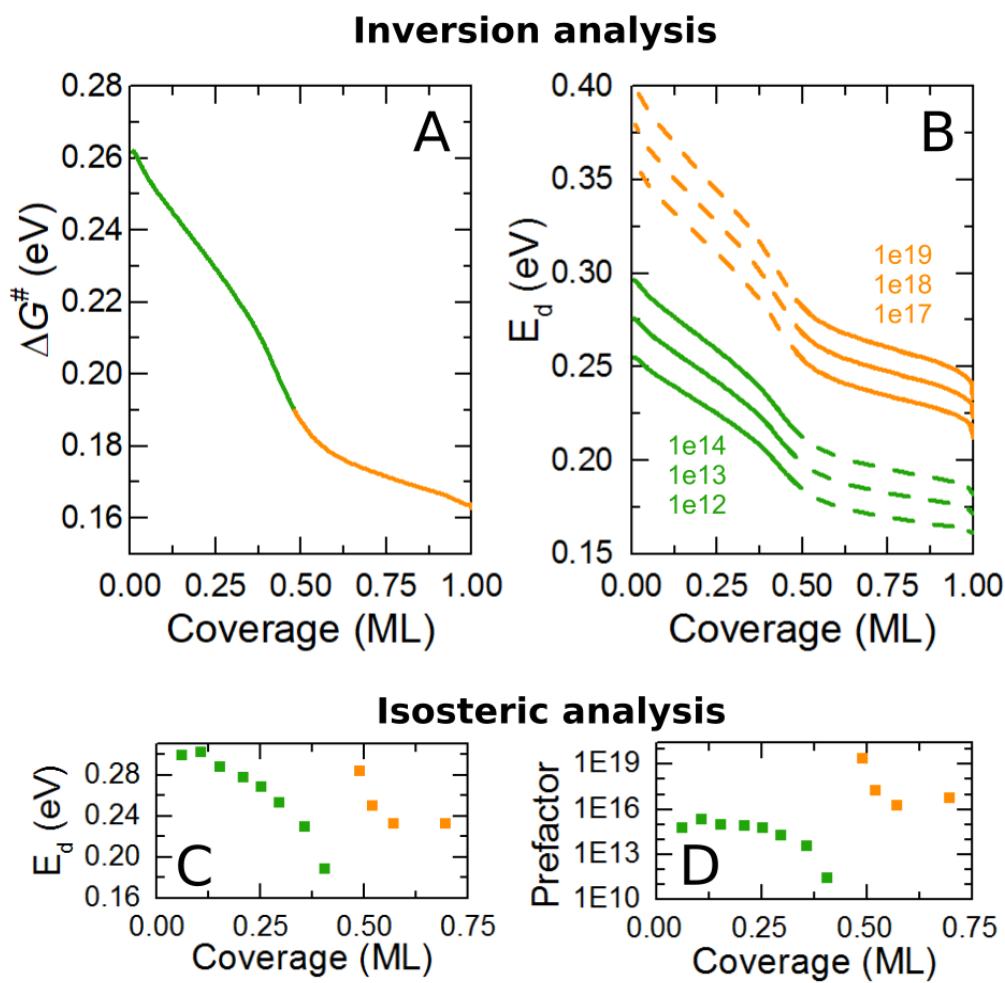

**Figure 3:** Analyses of the TPD data for CO on $Fe_3O_4$(001) shown in Fig. 2 A. **(A)** Inversion analysis showing $\Delta G^{\#}$ versus CO coverage in ML. **(B)** Inversion analysis, activation energy of desorption ($E_d$) vs. coverage in ML. Each curve corresponds to a different pre-exponential factor, which is assumed constant over the full range. **(C)** Isosteric analysis, activation energy of desorption vs. coverage. **(D)** Isosteric analysis, prefactor vs. coverage

The Polanyi-Wigner equation describing the kinetics of the desorption process is typically written as

$$-\frac{d\theta}{dt} = \nu\theta\exp\left(-\frac{E_d}{k_BT}\right) = -\frac{d\theta}{dT}\beta \qquad (1)$$

where $-\frac{d\theta}{dT}$ is the desorption rate, $\nu$ is the pre-exponential factor, $\theta$ is the coverage, $E_d$ is the activation energy of desorption, $k_B$ is the Boltzmann constant, $T$ is the temperature and $\beta$ is the heating rate. According to the transition state theory, the desorption equation can be written as

$$-\frac{d\theta}{dt} = \frac{k_BT}{h}\theta\exp\left(-\frac{\Delta G^{\#}}{k_BT}\right) \qquad (2)$$

where $\Delta G^{\#}$ is difference of the Gibbs free energy of a molecule in the transition state and a molecule in the adsorbed state, and $h$ is the Planck constant. Rearranging for $\Delta G^{\#}$, one obtains

$$\Delta G^{\#} = -k_bT\ \ln\left(-\frac{d\theta}{dt}\frac{1}{\theta}\frac{h}{k_bT}\right). \qquad (3)$$

Figure 3A shows $\Delta G^{\#}$, obtained from equation (3) using the TPD data acquired for an initial coverage of 1 ML. Two regions are clearly distinguishable: Coverages in the range 0-0.5 ML (0-2 CO molecules/surface unit cell) show a steeper decrease of $\Delta G^{\#}$ than coverages in the 0.5-1 ML range (2-4 CO molecules/surface unit cell).

To obtain the activation energy of desorption $E_d$ (aka. desorption energy), we have to separate the entropic and enthalpic parts of $\Delta G^{\#} = \Delta H^{\#} - T\Delta S^{\#}$, where $\Delta H^{\#}$ and $\Delta S^{\#}$ are the difference of the enthalpy and entropy of a molecule in the adsorbed state, respectively. We approximate $\Delta H^{\#}$ as $E_d$, which is accurate within a few meV (neglecting a $1/2\ k_BT$ term[25]). The prefactor $\nu$ is then given by

$$\nu = \frac{k_BT}{h}\exp\left(\frac{\Delta S^{\#}}{k_B}\right). \qquad (4)$$

To separately determine $E_d$ and $\nu$, we use the inversion procedure by Tait et al.[26], which means that we simulate the coverage-dependent TPD spectra assuming the Polanyi-Wigner equation (1) a constant prefactor $\nu$. The optimum prefactor is determined by the best fit between the simulated and experimental curves.

After performing such an analysis, it became clear that the assumption of a global prefactor cannot produce acceptable agreement over the whole coverage range. Since the two coverage regimes (green and orange curves in Fig. 2) are relatively well separated, we performed an independent inversion analysis to determine the optimum prefactor/$E_d$ combination. This is straightforward for the low coverage 0-0.5 ML regime; the TPD data for 0.5 ML is inverted and used to simulate the lower coverages only. We find the best agreement for $\nu=10^{13\pm1}$ s$^{-1}$, which yields $E_d$=0.275±0.02 eV in the low-coverage limit. The prefactor of $10^{13}$ is typical for a mobile weakly bound molecule with similar degrees of freedom to a gas phase molecule[25].

For analyzing the high-coverage regime (2 − 4 CO molecules/surface unit cell), we only use the $E_d(\theta)$ data in the range 0.56 − 1.00 ML. We obtain best agreement for $\nu = 10^{18\pm1}$ s$^{-1}$, which is close to the upper limit for the prefactors reported previously, and corresponds to a CO molecule with highly restricted degrees of freedom [27]. The weakness of this approach is that it still assumes a coverage-independent pre-exponential factor within each coverage regime, and assumes a discontinuity at 0.5 ML. Later, we will discuss how this can be understood in terms of a surface phase transition, with the discontinuity related to a entropy-enthalpy compensation effect [28].

For comparison, we performed the so-called isosteric analysis (also termed the "complete" analysis in the Ref. [29]), which allows to extract the desorption parameters without an initial assumption. Here, the values for desorption energy and prefactor are calculated from the set of Arrhenius plots of $\ln(-\frac{d\theta}{dt})$ vs $\frac{1}{T}$ constructed from the set of desorption curves of different initial coverages for a given value of coverage $\theta$. The desorption energy is obtained from the slope of the Arrhenius plot and the pre-exponential factor from the offset of the curve.

The results of the analysis are given in the Figs. 3 C and D. For the lowest coverages, we see that the values of desorption energy ($\approx$ 0.28 eV) and the pre-exponential factor ($\approx 10^{13}$ s$^{-1}$) match the conclusion obtained by the inversion analysis. The desorption energy and the pre-exponential factor decrease up to the coverage of ≈0.5 ML. After a discontinuity in the desorption energy and the prefactor for coverages at ≈0.5 ML, the values correspond to the higher-coverage TPD peak. There again, values of the desorption energy are close to the values obtained by the inversion method. The pre-exponential factor shows a similar trend to the desorption energy and the higher values for the higher coverages are in line with the results of the inversion analysis.

### 3.3 Scanning Tunneling Microscopy

In Figure 4 we show two STM images acquired at the same spot of the $Fe_3O_4(001)$-($\sqrt{2}\times\sqrt{2}$)R45° surface before and during exposure to $2\times10^{-9}$ mbar CO. The measurement was conducted using liquid nitrogen as the cryogen (sample temperature ≈86 K) because $Fe_3O_4$ is too insulating at liquid He temperatures to preform STM. The as-prepared surface shows the rows of surface $Fe^{3+}$ cations, which rotate by 90° at a monolayer step edge. Bright spots on these rows correspond to surface defects including an antiphase boundary (ADPB)[21], surface hydroxyls (OH)[22] and Fe adatoms ($Fe_{ad}$)[23], all marked in Fig. 4A. During exposure to CO, new bright scratchy features on some of the defect sites appear that are attributed to CO adsorbing on the surface defect sites. After prolonged exposure to CO, these features became obscured by rapidly diffusing adsorbates, and STM imaging became difficult (not shown). These observations are in agreement with the TPD experiments since the measuring temperature (86 K) is below the desorption temperature at defects. At this temperature, the CO molecules adsorbed at the regular lattice sites are mobile and are constantly ad- and desorbing from the sample.

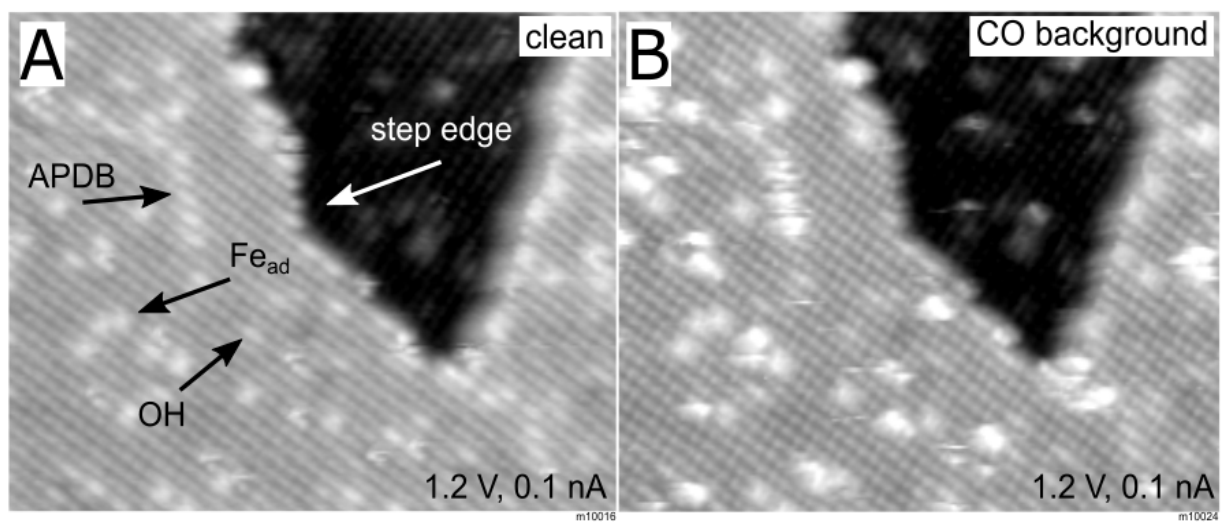

**Figure 4:** STM images (25 x 20 nm$^2$) of the as-prepared $Fe_3O_4(001)$ surface in UHV (A) and during exposure to CO (p = $2x10^{-9}$ mbar) (B). The measurements were conducted at 86 K. Marked in the clean surface image are typical defects visible in STM, namely step edges, surface hydroxyls (OHs), anti-phase domain boundaries (APDBs), and iron ($Fe_{ad}$).

### 3.4 UHV X-ray Photoelectron Spectroscopy

Figure 5 contains XPS spectra obtained from the CO/Fe$_3$O$_4$(001) system measured at 45 K for coverages of 0.43 ML and 1 ML. The O 1$s$ peak from the substrate is measured at 530.1 eV, consistent with prior studies. The peak from the adsorbed CO occurs at 536.6 V for 0.43 ML, and shifts to 536.4 eV at 1 ML coverage. The C 1$s$ peak from adsorbed CO is located at 290.6 eV for 0.43 ML, and shifts slightly to 290.3 eV for 1 ML. Both the O 1$s$ and C 1$s$ peaks appear slightly asymmetric towards the high-binding energy side. In general, the positions of the CO peaks are consistent with those observed for physisorbed CO on other metal oxide systems[11]. The slightly asymmetric shape is similar for both shown coverages of ~1.8 CO/u.c and ~4.3 CO/u.c. Therefore we don't expect it to be caused by separate peaks related to multiple adosrpiton sites. The asymmetric shape of the CO C 1$s$ and O 1$s$ peaks might be related to the vibrational broadening of the photoelectron peak[30] and/or to screening effects by the substrate[31], but to date there is no study addressing the photoelectron peak shapes of weakly bound molecules adsorbed on oxide surfaces. No change in the substrate related peaks, i.e., O 1$s$ or Fe 2$p$, is observed at any coverage (Fe 2$p$ not shown here).

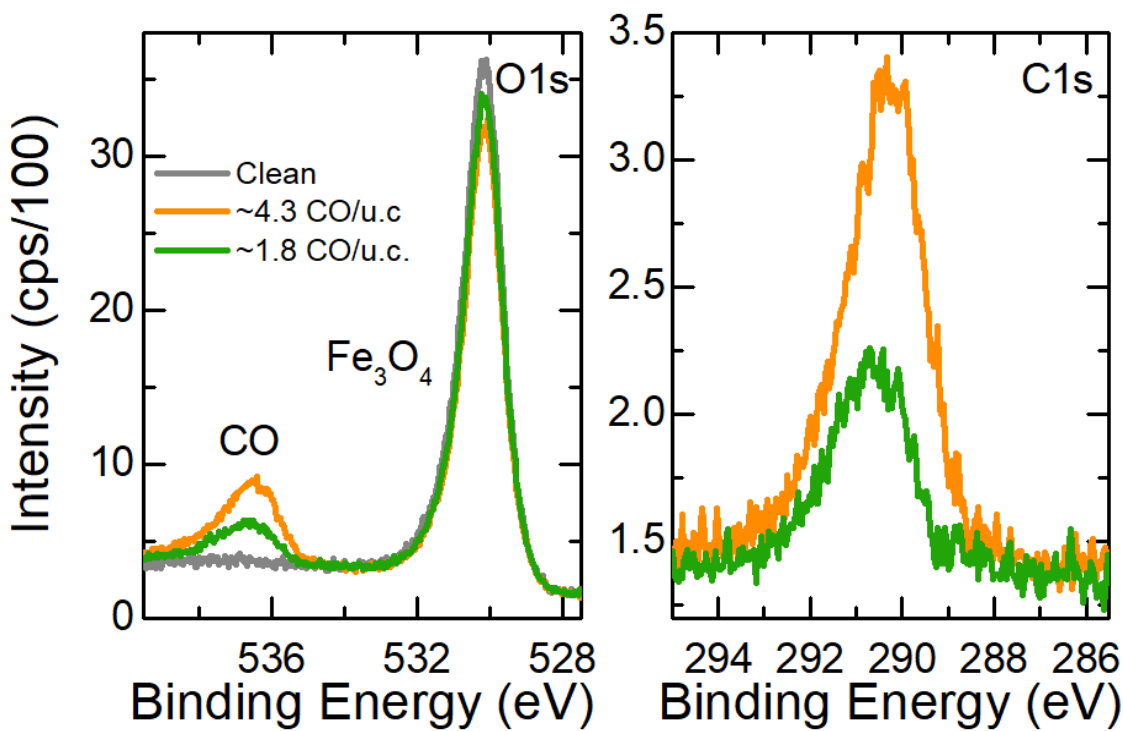

**Figure 5:** XPS spectra (monochromatic Al Kα) of different coverages of CO adsorbed on Fe$_3$O$_4$(001) at 45 K, measured under grazing emission (80° from the surface normal). **(A)** O1s region, **(B)** C1s region

## 3.5 Near Ambient Pressure X-ray Photoelectron Spectroscopy

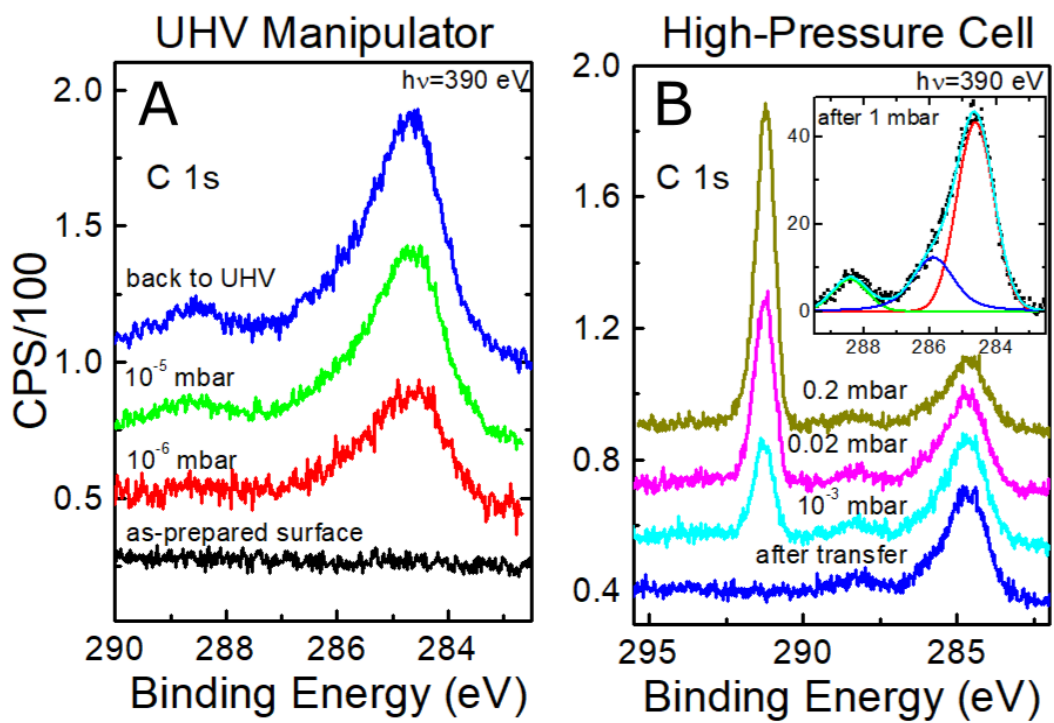

**Figure 6**: XPS measurements of the CO/Fe$_3$O$_4$(001) system performed using the NAP system at MAXlab at room temperature. **(A)** "UHV manipulator": The clean surface was measured following *in-situ* preparation, and then during exposure to CO partial pressures of $10^{-6}$ mbar and $10^{-5}$ mbar. The blue curve shows the C1$s$ region after the gas was pumped away before inserting the sample to the high-pressure cell. **(B)** "High-pressure cell": Spectra taken during exposure to CO partial pressures up to 0.2 mbar show no significant difference to those under UHV conditions and shown in (A), save for small intensity variations and the appearance of a gas phase CO peak at 291.2 eV. The final measurement (inset) was performed in UHV after exposure to the highest CO pressure (1 mbar). Peak fitting suggests three main contributions to the spectrum, located at 284.6 eV, 285.9 eV and 288.4 eV, respectively.

Although the UHV experiments suggest all CO is desorbed from the Fe$_3$O$_4$(001) surface well below room temperature, we performed NAP-XPS experiments at the I511 beamline at MAX IV Laboratory (see Fig. 6). The possibility that adsorption and surface reactions could be different at very low pressures and at those approaching values in the ambient is generally referred to as the "pressure gap" in surface science. The $Fe_3O_4$(001) single crystal prepared *in-situ* at the beamline end station exhibited a sharp ($\sqrt{2}\times\sqrt{2}$)R45° LEED pattern, as well as the expected Fe 2$p$ and O 1$s$ line shapes (not shown). No signal was present in the C 1$s$ region (black curve), but upon exposure to a partial pressure of $10^{-6}$ mbar CO an asymmetric peak emerged at 284.6 eV. The intensity of the peak grew slightly when the CO pressure was raised to $10^{-5}$ mbar, and a small peak emerged at 288.4 eV. Subsequently the sample was moved to the high-pressure cell. No significant change in the appearance of the C 1$s$ region occurred following the removal of the gas, nor following the transfer between chambers. The sample was then exposed to higher CO partial pressures between $10^{-2}$ mbar and 0.2 mbar, with the only major change being the emergence of a peak for gas-phase CO at 291.2 eV, and a slight decrease in the overall intensity of the signal emerging from the sample surface. At 1 mbar, the gas phase CO signal became very intense and the signal from the surface could not be resolved. Nevertheless, once the gas was pumped from the chamber, the C 1$s$ signal showed again no change from what was observed at $10^{-5}$ mbar. The final spectrum could be fit well using a Shirley background and 3 Voigt peaks centered at 284.6 eV, 285.9 eV, and 288.4 eV. These components have their origin most likely in the (hydrocarbon) impurities in the CO gas, and are typical for carbonaceous deposits often found on samples exposed to high gas pressures/atmosphere. At no point during or after the exposure of the surface to elevated pressure of CO did we observe any changes in the substrate-related Fe 2$p$ signal (Figure S3 in the supporting information). This suggests that CO induces neither surface reduction nor iron carbide formation at room temperature in this pressure regime.

It is important to note that the gas-phase CO peak (~291.2 eV) is shifted from the peak of the physisorbed CO observed at low temperature by ~ 0.5 eV – 0.8 eV to higher binding energies. Since the desorption energy of CO is small compared to the chemical potential of the CO gas at the applied pressures at 300 K, occupation of the weakly bound, physisorbed state is too low to be seen by photoemission experiments at these conditions.

4. Discussion

The sticking coefficient for CO on $Fe_3O_4$(001) at low temperatures is high (above 0.8) with no strong coverage dependence, which suggests there is no significant barrier for CO adsorption on this surface. Thus, we can compare the activation energy for desorption determined from TPD directly to the adsorption energy calculated by DFT[25]. Under UHV conditions, low coverages of CO desorb from the regular terrace already by 105 K, which corresponds to an activation energy for desorption of ≈0.28 eV for an isolated molecule according to the inversion analysis. The DFT+U calculations of Xue et al.[14] predict an adsorption energy of 0.24 eV for a CO molecule bound through the C atom to a surface $Fe^{3+}$ cation. Given the periodic boundary conditions employed in these calculations, this corresponds to a coverage of 1 CO molecule per (√2×√2)R45° unit cell. At this coverage (0.25 ML), the inversion analysis of our TPD data determines an activation energy for desorption of ≈0.24 eV. In contrast to the predictions of Xue et al.[14], we observe a decrease in CO binding energy with coverage. We find no evidence for C-C bond formation or reduction of the surface by CO at room temperature. The latter process likely becomes possible at elevated temperatures, because, as mentioned in the introduction, heating in CO is how magnetite-containing minerals are reduced to iron ore. Previously, we have seen that lattice O can be extracted from this surface in the presence of Pt nanoparticles by exposure to CO at 520 K[32].

In the low coverage regime (0 − 0.5 ML), the CO TPD peak shifts rapidly to lower temperatures. This phenomenon has been observed previously and attributed to repulsion between adsorbed molecules in a mobile 2D gas [9-10], which increases as more molecules are packed in. The effect of nearest-neighbour dipole repulsion has been extensively modeled for CO on metal surfaces, and strong interactions can produce two distinct TPD peaks in the first ML, similar to what we observe in Figure 2A[19]. However, the discontinuity in the quantitative analysis of the TPD data at two molecules per unit cell indicates a phase transition. We propose that below two molecules per unit cell, the adsorbed CO is a 2D gas, consistent with the $\nu$ of $10^{13}$/s. Above this coverage, the occupation of nearest neighbor $Fe^{3+}$ sites cannot be avoided, and the system becomes ordered. Both $E_d$ and $\nu$ are different for desorption from the condensed phase. Note that the Gibbs free energy of desorption is (and must be) continuous through such a phase transition [27] since the two phases coexist at the coverages around the phase transition at ≈ 0.5 ML. A discontinuous increase in $E_d$ is compensated by a corresponding increase in entropy, and, hence, $\nu$. This phenomenon is known as enthalpy-entropy compensation[27]. According to transition state theory, pre-exponential factors as high as $10^{18}$ can occur if there is a large entropy change between the adsorbed and transition states, i.e., if a molecule desorbs from a (locally) ordered state in which translation, vibrational, and/or rotational degrees of freedom are hindered. A phase transition near 0.5 ML can also explain the cusps in the isosteric analysis (Fig. 3C, D): As mentioned above, this analysis is based on fitting Arrhenius plots, which results in unphysical values if the fit includes data across a phase transition.

These results show that the quantitative TPD analysis as applied here has only limited applicability for desorption systems with complicated desorption kinetics. Although the observed dependence of the kinetic parameters can be qualitatively rationalized by the lateral repulsion between the CO molecules and the existence of two phases, addressing the kinetic aspects of the complex desorption processes without more sophisticated theoretical models is beyond the scope of this paper.

At this point we note the similarity between the TPD data in Figure 2A and those obtained for the stoichiometric $TiO_2(110)$ surface[9-10]. TPD from that system exhibits a similar two-peak structure for CO in the first monolayer, with a transition close to 0.5 ML coverage. Based on IRAS data, Petrik and Kimmel [10] demonstrated that both peaks originate from CO molecules bound in a perpendicular orientation at surface Ti cations, in line with our explanation for $Fe_3O_4(001)$. The binding of an isolated molecule appears stronger on $TiO_2(110)$ than on $Fe_3O_4(001)$, with desorption occurring at ≈ 150 K in TPD. Dohnalek et al.[9] determined the activation energy for desorption to be 0.41-0.46 eV (using a preexponential factor of $10^{13}$ $s^{-1}$ via an inversion analysis, whereas Petrik and Kimmel [10] determined 0.36 eV using a Redhead analysis assuming the same prefactor. Interestingly, in both cases the desorption from the close-packed state occurs at 65 K, most likely because the arrangement and distance of the binding sites is similar between these two surfaces, and CO-CO interactions begin to dominate at this adsorbate density. On $Fe_3O_4(001)$, the cations are arranged in rows with a periodicity of 3 Å and a spacing of 6 Å (Figure 1 C). On $TiO_2(110)$, the periodicity is 2.9 Å and the inter-row spacing is 6.5 Å.

Next, we discuss our results in the context of CO adsorption as a probe of the acidic sites on metal-oxide surfaces. One might expect that that CO would interact more strongly with $Fe^{3+}$ sites than $Fe^{2+}$ sites on $Fe_3O_4$ surfaces, given that the $Fe^{3+}$ ion is a stronger Lewis acid. However, the SCV reconstruction is oxidized with respect to bulk $Fe_3O_4$, and contains only $Fe^{3+}$ cations in the surface layer[15]. This gives rise to the main TPD peaks in Figure 2A. In addition to desorption from the main terraces, we observe three small peaks at higher temperatures, which we assigned to "surface defects" above. The main surface defects on $Fe_3O_4(001)$ are step edges, APDBs in the surface reconstruction[23], Fe adatoms[23], excess Fe atoms incorporated in the subsurface layer[23], and surface hydroxyl groups[22, 33]. While each of these defects appears quite different, the latter three have in common that they induce charge transfer into neighboring surface Fe atoms, which appear brighter in both empty and filled states STM images[1], suggesting increased DOS close to $E_F$ in comparison to the regular $Fe^{3+}$ cations. Indeed, the modified DOS at the surface Fe atoms due to hydroxyl groups was recently measured by scanning tunneling spectroscopy [34]. This effect results in stronger CO binding because both charge transfer from the CO 5σ orbital to the cation (requiring available empty states), and back donation of the surface electrons from the surface to the CO 2π* orbital (requiring filled states near $E_F$) contribute to the bond. This is in line with the generally observed fact that CO binds more strongly to the metal cations with the oxidation state closer to neutral state[35].

Finally, we turn our attention to the synchrotron-based NAP-XPS experiments. As mentioned above, we rationalize the lack of physisorbed CO in the spectra as the direct result of the weak CO binding observed in UHV. Given the activation energy for desorption of 0.3 eV, the instantaneous CO coverage at a gas pressure of 0.2 mbar will only be of the order $1\times10^{-8}$ ML; too small to detect by XPS. Nevertheless, we do observe peaks in the C 1s region that were not observed in the low temperature XPS experiments. The peak at 288.4 eV is very close to that measured for formate on this surface[36], and it is certainly possible that CO reacts with surface OH groups, or with water from the residual gas to form this species given that $Fe_3O_4$ is a well-known water-gas shift catalyst. At this stage, however, we cannot rule out the alternative possibility that this peak is due to carbonate species. Finally, the peaks at 284.6 eV and 285.9 eV are typical for O-C-O and C-C bonds, most likely due to the unintended exposure of the surface to hydrocarbons in the high-pressure environment.

## 5    Conclusion

The $Fe_3O_4$(001) surface shows weak interaction with CO, desorbing from the $Fe^{3+}$ regular lattice sites in the temperature range 55 K – 105 K. The adsorption energy of an isolated molecule calculated from the TPD experiments is 0.28 eV. Above a coverage of two molecules per unit cell, the system undergoes a phase transition from a 2D gas to an ordered phase. The strongest CO adsorption occurs at $Fe^{2+}$ related defects, with distinct peaks observed at 120 K and 220 K. Near-ambient-pressure experiments reveal no sign of the surface reduction or C-C bond formation, but do hint at a possible reaction between physisorbed CO and water from the residual gas.

**Supporting Information is available:** Additional TPD experiments allow to identify the origin of defect states, and reveal the compression of the monolayer prior to multilayer formation. Supplementary AP-XPS data show only minor changes to the Fe2p and O1s spectra following high pressure CO exposure.

**Acknowledgements**

The authors gratefully acknowledge funding through projects from the Austrian Science Fund FWF (START-Prize Y 847-N20 (JH & GSP); Special Research Project 'Functional Surfaces and Interfaces, FOXSI F45 (MS & UD)), the European Research Council (UD & ZN: ERC-2011-ADG_20110209 Advanced Grant 'OxideSurfaces') and the Doctoral College TU-D

**TOC Graphic**

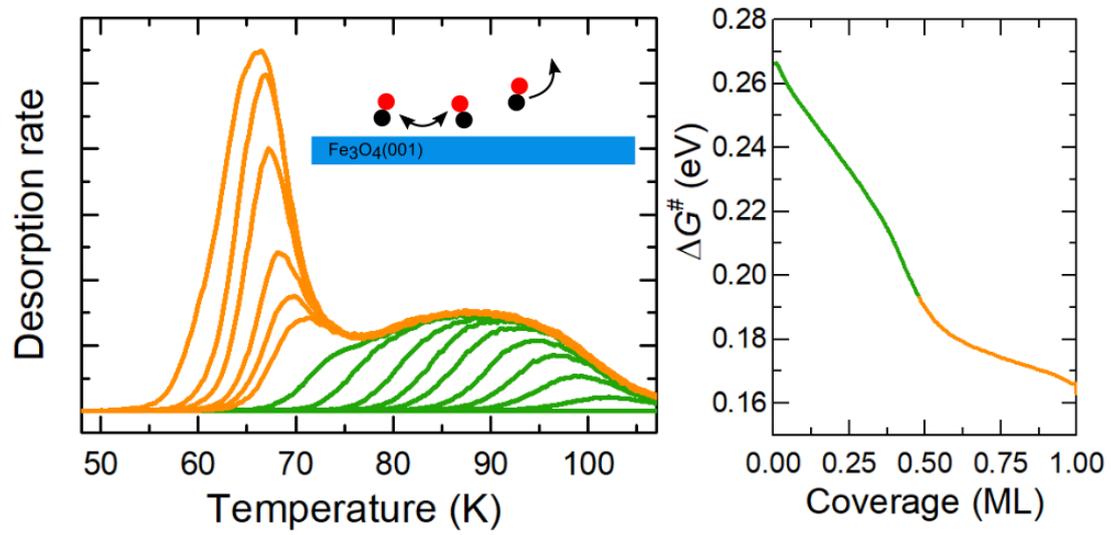

**Author list**

Mr. **Jan Hulva**
 - e-mail: hulva@iap.tuwien.ac.at
 - current association: Institute of Applied Physics, TU Wien
 - address: Wiedner Hauptstraße 8-10/134, 1040 Wien, Austria
 - telephone: +43 1-58801-13438

Mr. **Zdeněk Jakub**
 - e-mail: jakub@iap.tuwien.ac.at
 - current association: Institute of Applied Physics, TU Wien
 - address: Wiedner Hauptstraße 8-10/134, 1040 Wien, Austria
 - telephone: +43 1-58801-13478

Dr. **Zbynek Novotny**
 - e-mail: novotny@physik.uzh.ch
 - current association: Department of Physics, University of Zürich
 - former association: Institute of Applied Physics, TU Wien
 - address: Winterthurerstrasse 190, 8057 Zürich, Switzerland
 - telephone: +41 44 635 6691

Mr. **Niclas Johansson**
 - e-mail: niclas.johansson@sljus.lu.se
 - current association: Division of Synchrotron Radiation Research, Lund University
 - address: Box 118, SE-221 00 Lund, Sweden
 - telephone: +46462223894

Dr. **Jan Knudsen**
 - e-mail: jan.knudsen@maxiv.lu.se
 - current association: Division of Synchrotron Radiation Research, Lund University,
       MAX IV Laboratory, Lund University
 - address: Box 118, SE-221 00 Lund, Sweden
 - telephone: +46462228283

prof. **Joachim Schnadt**
 - e-mail: joachim.schnadt@sljus.lu.se
 - current association: Division of Synchrotron Radiation Research, Lund University,
 - address: Box 118, SE-221 00 Lund, Sweden
 - telephone: +46462223925

prof. **Michael Schmid**
 - e-mail: schmid@iap.tuwien.ac.at
 - current association: Institute of Applied Physics, TU Wien

  - address: Wiedner Hauptstraße 8-10/134, 1040 Wien, Austria
  - telephone: +43 1-58801-13452

prof. **Gareth S. Parkinson\* (corresponding author)**
  - e-mail: parkinson@iap.tuwien.ac.at
  - current association: Institute of Applied Physics, TU Wien
  - address: Wiedner Hauptstraße 8-10/134, 1040 Wien, Austria
  - telephone: +43 1-58801-13473